\documentclass[12pt]{article}
\usepackage[]{graphicx,rotating,times,pdfsync,epstopdf}

\begin{document}
\begin{center}
{\Large\bf Search for WIMPs in liquid argon\footnote{Proceedings of the WIN'11 Conference, Cape Town (2011)}}\\
\bigskip

C. AMSLER\footnote{E-mail: claude.amsler@cern.ch, http://amsler.web.cern.ch/amsler/}\\
Physik-Institut der Universit\"at Z\"urich\\
CH-8057 Z\"urich, Switzerland\\

\end{center}

\noindent
\underline{\bf Abstract:}\\
{\it Our group from the University of Zurich is performing R\&D work towards the design of a large liquid argon detector to detect Weakly Interacting Massive Particles (WIMPs). This project is developed within the DARWIN Collaboration funded by ASPERA to prepare a proposal for the next generation of WIMP searches using noble liquids.  We are performing R\&D to detect the VUV light from recoiling argon nuclei. Results obtained with one ton of liquid argon (ArDM prototype) and prospects using a monoenergetic neutron source are discussed.}

\section{Introduction}
From astronomical observations it is known that most of the gravitational mass in the universe does not couple to electromagnetic radiation and is therefore not directly observable with telescopes. Recent measurements  lead to the conclusion that about 72\% of the mass density is due to dark energy, 23\% to dark matter (DM), and only 5\%  to baryons \cite{Drees}. The contribution from neutrinos is less than 1\%. Baryons are mostly located in the intergalactic gases with only about 10\% contributing to star masses. Roughly 15\% of the DM may  be due to MACHOS or cold molecular clouds. Although exotic mecanisms  (such as modifications of Newton's gravitational law \cite{Mond}) that do not require DM  have been postulated, the natural assumption is that DM indeed exists as some form of elementary particles, axions or Weakly Interacting Massive Particles (WIMPs), or possibly as primordial black holes (having formed before nucleosynthesis).  The direct laboratory observation of dark matter (DM) is  one of the most pressing issues in Particle Physics, which can be addressed at the LHC and with non-accelerator experiments. 

\begin{figure}[htb]
\parbox{80mm}{\mbox{
\includegraphics[width=80mm]{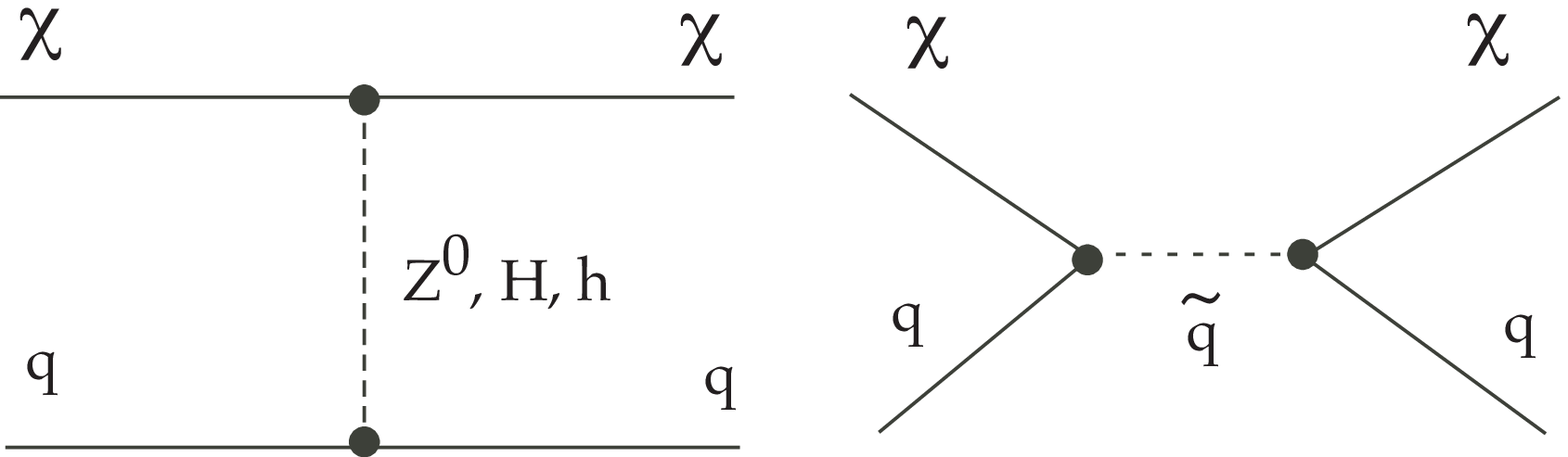}
}\centering}\hfill
\parbox{80mm}{\mbox{
\includegraphics[width=80mm]{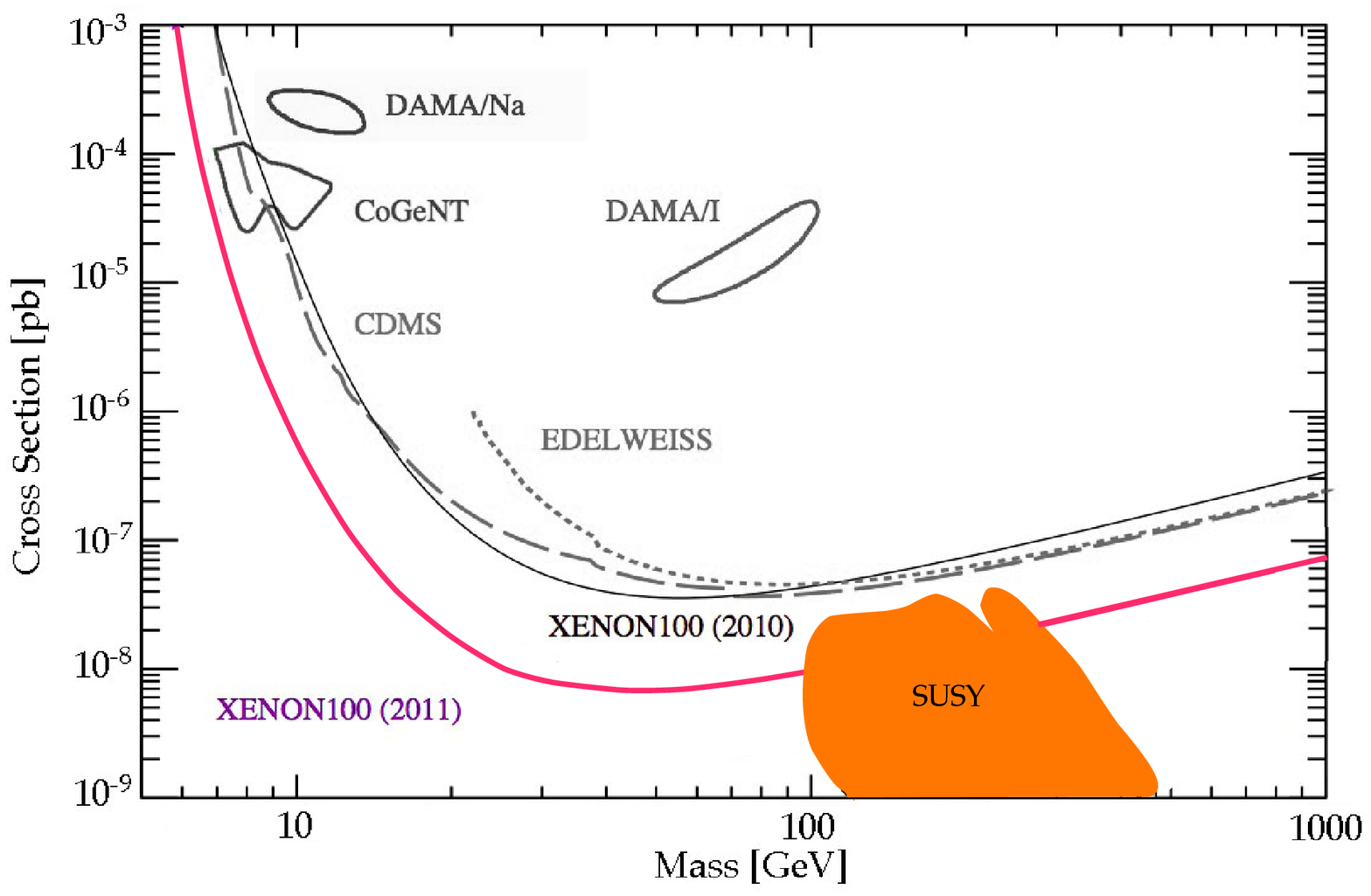}
}\centering} \caption[]{\it Left: Feynman graphs of $\chi\equiv\tilde{\chi}_1^0$ interactions with quarks in the nucleon. Right: WIMP-nucleon cross section as a function of WIMP mass showing the most stringent upper limit (90\% CL) from XENON-100 (adapted from ref.  \cite{Aprile}). Theoretical predictions from supersymmetry \cite{Buch} are also shown.
\label{Neutralino}}
\end{figure}

Because DM has survived since the birth of the universe it has to be stable and only weakly interacting. As mentioned already, the most popular candidates for DM are axions and WIMPs. Heavy neutrinos would  qualify as WIMPs, although there is no obvious reasons as to why they should be stable. The lightest superparticle in SUSY models conserving $R$-parity is the most popular WIMP candidate, the spin 1/2 neutralino $\tilde{\chi}_1^0$ with  mass  in the 10 GeV to 10 TeV range. $R$-parity conservation ensures that the $\tilde{\chi}_1^0$  is stable. Also, the $\tilde{\chi}_1^0$ cannot transform  into other SUSY particles when interacting with matter, due to  its low mass. At the LHC the $\tilde{\chi}_1^0$ will therefore manifest itself by a large missing energy. On the other hand, the $\tilde{\chi}_1^0$
can scatter e.g. on constituent quarks in nucleons or nuclei (Fig. \ref{Neutralino}, left), leading to nuclear recoils in the range of 1 -- 100 keV. Non-accelerator laboratory searches are all based on the detection of such nuclear recoils. The scattering cross section on nucleons is tiny, in the range 10$^{-5}$ to 10$^{-12}$ pb, comparable to that for  neutrino interactions. The differential cross section decreases exponentially with recoil  energy (see Fig. \ref{Formfactor} below) which makes WIMP detection difficult due to the low energy background. Hence massive detectors with low detection thresholds are needed, among them cryogenic ones using liquid argon (LAr) or liquid xenon (LXe).  

WIMPs can annihilate in the galactic halo, leading to the production of e.g. positrons, antiprotons, neutrinos or photons. For instance, an excess of $\gamma$-rays consistent with 80 GeV WIMPS has been reported in a re-analysis of EGRET data \cite{deBoer}. Satellite experiments such as GLAST, AMS2, PAMELA, \v{C}erenkov detectors such as MAGIC and VERITAS, or large neutrino telescopes such as ICE CUBE, might reveal signals from WIMP annihilations \cite{Drees}. An annual modulation (6.3$\sigma$ effect) was reported by the DAMA experiment with 100 kg of NaI detectors \cite{DAMA}, and more recently confirmed  by the same group (8.2$\sigma$ effect with 250 kg of NaI). The signal was, however, not observed by other  experiments. In particular, the spin-dependent signal was recently ruled out by COUPP \cite{Behnke}. 

Fig.  \ref{Neutralino} (right) shows upper limits for WIMPs from recent experiments. Since one does not know precisely the cross section, one usually shows the limits in a two dimensional plot  cross section vs. mass. For low masses the sensitivity decreases,  due to the low recoil energy and the detection threshold. For high masses the loss of sensitivity is due to the diminishing WIMP flux. The best upper limits are obtained for a WIMP mass comparable to the atomic number $A$ of the target. The spin independent rate provides the current best upper limits for WIMPs and is also the relevant interaction for our developments with liquid argon since   $^{40}$Ar is a spin zero nucleus. The XENON-100 \cite{Aprile} and CDMS \cite{Ahmed} experiments have produced the best upper limits so far, the former even reaching  7 $\times$ 10$^{-9}$ pb for masses around 50 GeV. The current upper limits now reach into the relevant region of theoretical predictions based on supersymmetry \cite{Buch}.

\begin{figure}[htb]
\parbox{80mm}{\mbox{
\includegraphics[width=60mm]{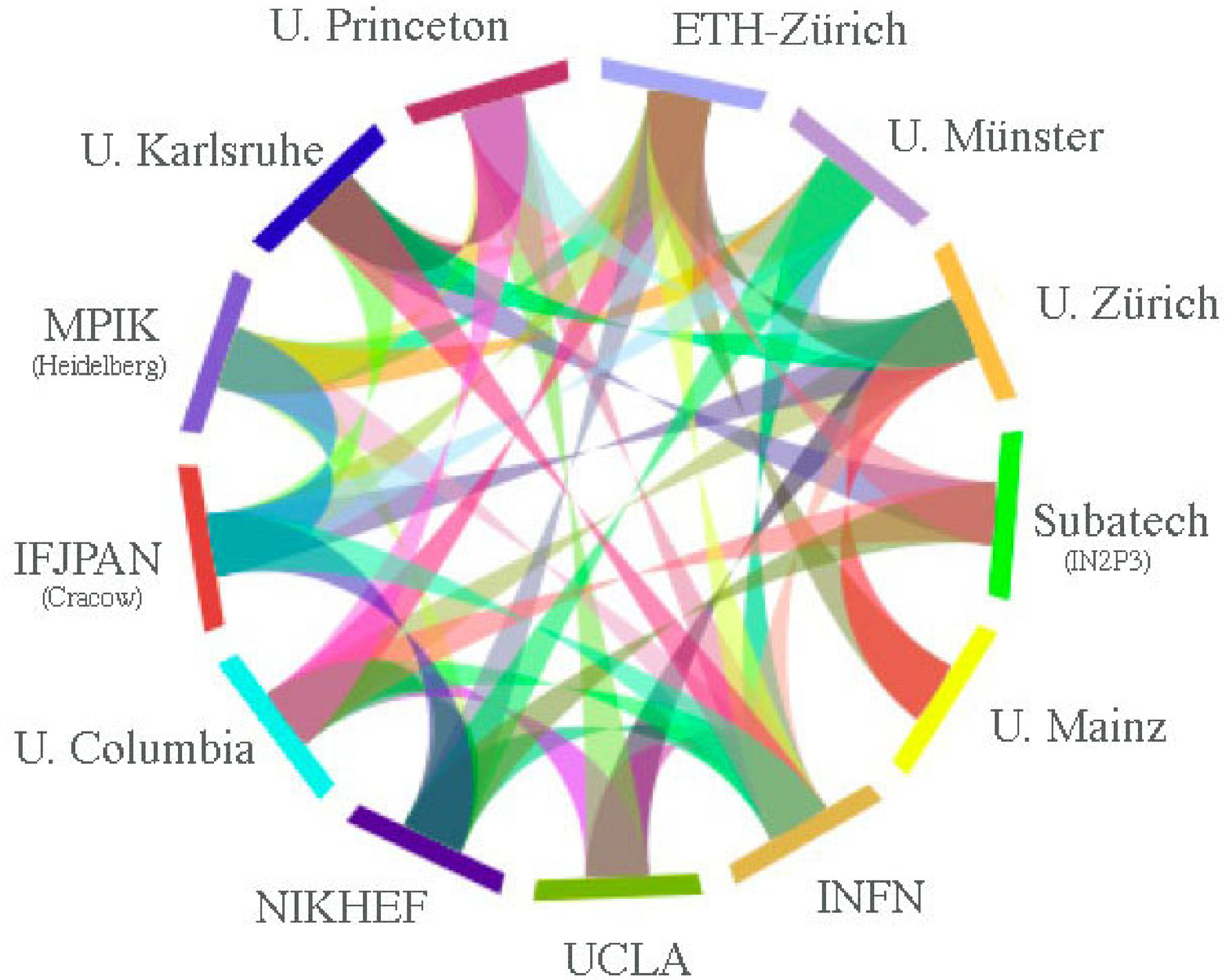}
}\centering}\hfill
\parbox{80mm}{\mbox{
\includegraphics[width=60mm]{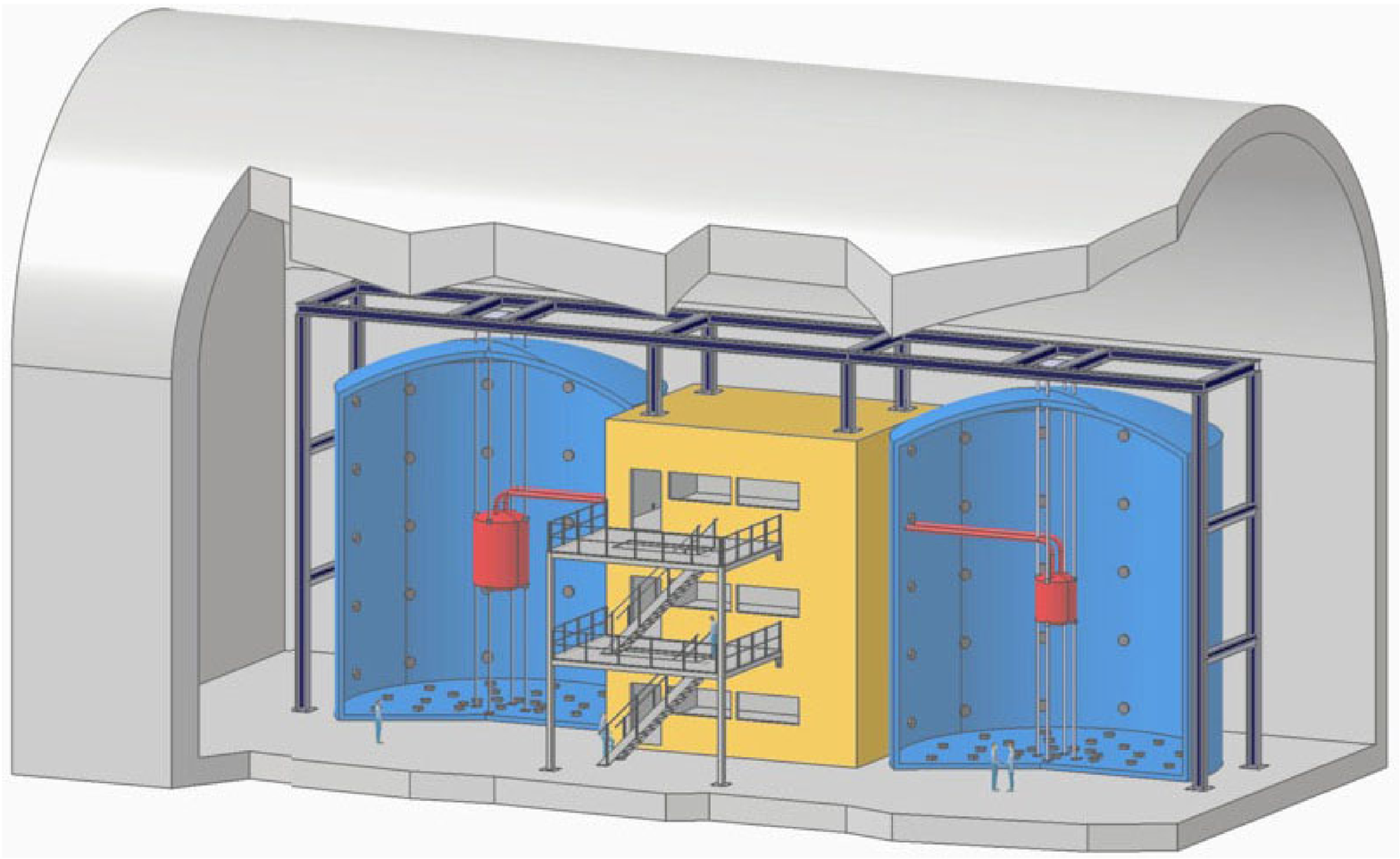}
}\centering} \caption[]{\it Left: Institutions participating in DARWIN. Right: Possible layout of a large underground detector using 8t of LAr and 5t of LXe, surrounded by water \v{C}erenkov veto shields \cite{Baudis}.
\label{LArLXe}}
\end{figure}

Most WIMP searches can be classified in two classes, (1) those using pure semiconductors and, (2) those using noble liquids such as LXe or LAr. In semiconductors operating at very low temperatures one measures the phonons and/or the ionization generated by recoiling nuclei. Representatives of this class are CDMS, CRESST, EDELWEISS. In noble liquids (ArDM, DEAP, WARP, XENON, ZEPLIN) one mesures the UV light and possibly also  the electric charge produced by the ionizing recoils. Sensitivites will have to be improved by at least two orders of magnitude in future projects. This requires massive detectors operated by large international collaborations.  Accordingly, ASPERA has funded two projects, EURECA \cite{EURECA} based on solid state detectors, and DARWIN \cite{DARWIN} using noble liquids. The latter consortium is to deliver the concept for a multiton LAr/LXe detector by 2013 which could become operational around 2016, reaching a final sensitivity of 10$^{-11}$ pb. Fig. \ref{LArLXe} shows the  possible layout of such a detector.

\section{Experimental sensitivities in noble liquids}
\begin{figure}[htb]
\parbox{160mm}{\mbox{
\includegraphics[width=80mm]{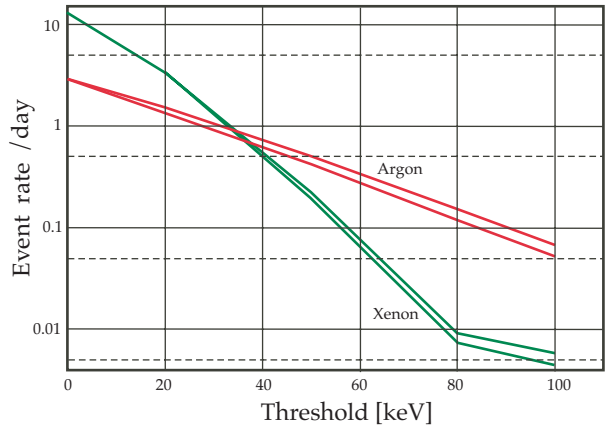}
}\centering}\hfill
\caption[]{\it Daily WIMP detection rate for a 1t detector as a function of detection threshold for LXe and LAr, assuming a WIMP mass of 100 GeV and a cross section of 10$^{-8}$  pb. The top and bottom curves show the limits of annual variations due to the rotation of the Earth around the Sun (after ref.  \cite{Chandra}).
\label{Formfactor}}
\end{figure}

One often assumes  that DM is distributed within a non-rotating spherical halo larger than the Milky Way. The velocity distribution is taken as Maxwell-Boltzmann with a dispersion of 220 km/s. The solar system rotates around the galactic center with a velocity of 230 km/s and the earth around the sun with a velocity of 30 km/s. Taking into account the 60$^\circ$ inclination of the ecliptic one then finds that the average velocity of WIMPs entering a terrestrial detector is  $\sim$245 km/s in June and $\sim$215 km/s in December. The differential scattering cross section on a nucleus is given by \cite{Jungman} 
\begin{equation}
\frac{d\sigma}{dq^2} = \frac{G_F^2 C F^2(q^2)}{v^2},
\label{crse}
\end{equation}
where $q^2$ is the momentum transfer to the nucleus, $G_F$  the Fermi coupling constant, $C$ a constant describing details of the interaction, $F$ a nuclear form factor and $v$ the WIMP velocity in the detector. The event rate in a detector can be calculated \cite{Chandra} by folding the cross section with the Boltzmann distribution, assuming a density of WIMPs of about 0.3 GeV/cm$^3$. The cross section consists of two components, a nuclear spin dependent part and a spin independent part. The former is larger on single nucleons, but constructive interference enhances the latter by a factor of $A^2$ for nuclei. Small WIMP velocities are favored by the cross section (eq. \ref{crse}) and the form factor suppresses large recoil energies. Hence the distribution falls off  (roughly exponentially) as a function  of nuclear recoil energy. This makes WIMP detection difficult due to the low energy background. Fig.  \ref{Formfactor}  shows  the recoil energy distribution in 1t of  liquid argon and liquid xenon detectors assuming a cross section of 10$^{-8}$  pb. Recoil energy spectra in LAr and LXe are different. These liquids are therefore complementary  in providing a crosscheck in case a  WIMP signal is observed.  With a  nuclear recoil energy threshold of  30 keV, a WIMP-nucleon cross section of 10$^{-8}$  pb would yield 1 event per day and per ton of LAr.  Even a 1t  LAr detector could improve significantly the current upper limits, provided that the threshold of 30 keV and sufficient background rejection can be achieved. Statistics permitting, the annual variation could even be used as  additional evidence for the detection of WIMPs.

\section{Light detection in a 1t liquid argon detector}
The ArDM Collaboration \cite{ArDM} uses the double-phase argon technique  pioneered by the WARP collaboration \cite{Calligarich}.  ArDM envisages a much larger and more ambitious detector  (1t of LAr). In contrast to WARP one  plans to detect both charge and luminescence from WIMP interactions. Charged particles lead to ionization and excitation of argon atoms, forming excimers with  the lowest singlet and the triplet excited states decaying by  VUV photon emission in a narrow band around 128  nm. The  singlet and the triplet states have  different decay times, respectively $\tau_{1}\simeq 5$ ns and $\tau_{2}\simeq 1.6$ $\mu$s in liquid argon (LAr) \cite{Hitachi}.  The slow component is very sensitive to the purity of the argon \cite{amsler1}. Heavily ionizing particles such as $\alpha$'s or nuclear recoils contribute mostly to the fast ($<$ 50 ns) decaying component, while the contribution of electrons and $\gamma$'s  to the slow component is larger.  For minimum ionizing projectiles, such as  electrons and $\gamma$'s, the component ratio $CR$ (fast to total intensity)  is $  \approx 0.3$, while for $\alpha$'s and nuclear recoils $CR \simeq 0.8$ \cite{Hitachi}. In addition, the ionization yield is much lower for nuclear than for minimum ionizing particles, due to quick recombination. Both features can be used in LAr to reduce background in WIMP searches. 

LAr has a density of 1.4 g/cm$^3$, a refractive index of 1.24 and boils at 87 K. LAr generates about 5 $\times$ 10$^4$ VUV photons/MeV for minimum ionizing particles (in the absence of electric field). LAr is cheap (1.5 CHF/$\ell$) compared to xenon, but has the disadvantage of containing the radioactive nuclide $^{39}$Ar ($\beta$ - emitter with a half life of 269 yrs and  565 keV endpoint energy). The activity of natural argon is  1 Bq/kg  which induces a background rate  of roughly  1 kHz in a 1t  detector \cite{Benetti}. 

\begin{figure}[htb]
\parbox{160mm}{\mbox{
\includegraphics[width=140mm]{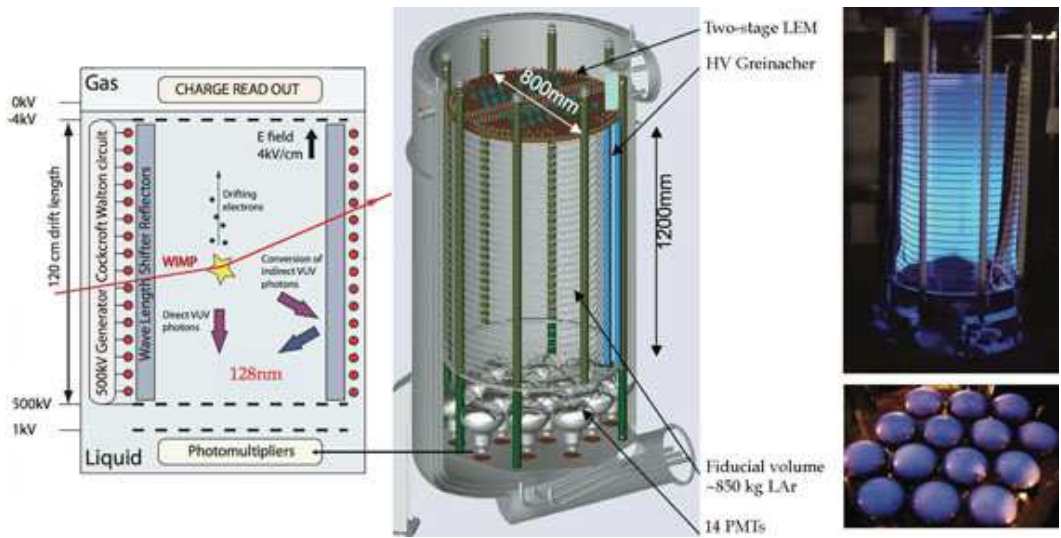}
}\centering}\hfill
\caption[]{\it Left: Detector concept. Middle: PMT array and HV divider chain. Right: WLS-foils and PMTs under UV illumination \cite{amsler2}.
\label{ArDMdetector}}
\end{figure}

Details on the ArDM detector can be found  in recent publications \cite{BocconeJINST,amsler2}. Briefly, the working principle  is as follows (Fig. \ref{ArDMdetector}, left): in LAr a  WIMP collision leading to 30 keV nuclear recoils produces about 400 VUV (128 nm) photons, together with a few free electrons. The latter are drifted in a strong vertical electric field and are detected in the gas phase by a large electron multiplier (LEM) above the surface of the liquid, while the VUV  scintillation light is shifted into blue light by a wavelength shifter (WLS) and detected by  cryogenic photomultipliers at the bottom of the vessel. Fig. \ref{ArDMdetector} (middle)  shows a sketch of the detector with the 400 kV HV divider. The detector and its recirculation and purification systems   are currently installed in building 182 at CERN for preliminary readout tests. The detector for the charge extracted from the liquid  into the gas phase with subsequent multiplication in a prototype LEM  still needs to be installed.  

\begin{figure}[htb]
\parbox{80mm}{\mbox{
\includegraphics[width=60mm]{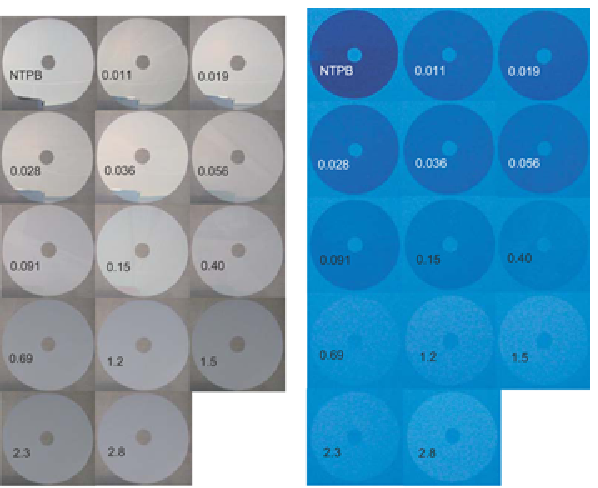}
}\centering}\hfill
\parbox{80mm}{\mbox{
\includegraphics[width=60mm]{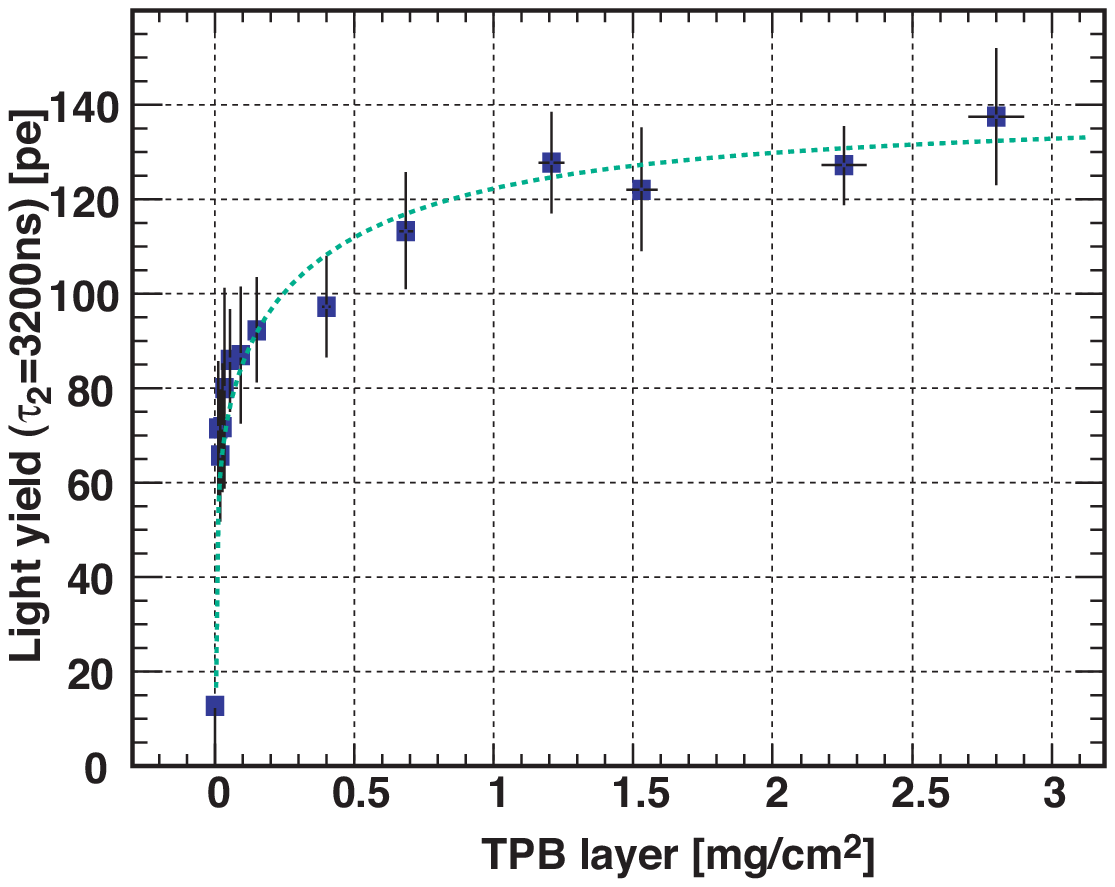}
}\centering} \caption[]{\it Left: disks of ESR-foils covered with TPB under ambient  and UV light. Right: Light yield in  photoelectrons (pe) as a function of WLS thickness. 
\label{disks}}
\end{figure}

At the University of Zurich we have designed and built the light readout system for the 1t LAr  detector. We measured the light yield dependence on the WLS thickness of evaporated specular reflectors foils. The measurements were made with a small cell containing  gaseous argon at atmospheric pressure. Two materials were selected, Tetratex and ESR-foils from 3M, and the optimum thickness of the TPB layer on the reflector was determined experimentally \cite{Cabrera}. The best conversion efficiency, uniformity and reproducibility were achieved by evaporating TPB on the reflector. Several disks of ESR foils (diameter 70\,mm) were covered with TPB layers of different thicknesses. The TPB thickness was determined by weighing the disks before and after evaporation. The increase in brightness with TPB thickness is apparent under UV illumination (300 nm)  (Fig. \ref{disks}, left). The response to 128 nm light was determined using scintillation light from gaseous argon and an $\alpha$-source.  The light yield of the slow component ($\tau_2$ = 3.2 $\mu$s in gas \cite{amsler1}) is shown in Fig. \ref{disks} (right) as a function of thickness. The data are consistent with a saturation of VUV conversion efficiency above 1 mg $\cdot$ cm$^{-2}$. Accordingly, fifteen Tetratex WLS sheets (120 $\times$ 25 cm$^2$) were coated  with 1 mg $\cdot$ cm$^{-2}$ TPB to cover the cylindrical side walls inside the electric field shapers (Fig. \ref{ArDMdetector}, middle). 

The  light detection system consists of fourteen 8'' hemispherical  photomultipliers (PMT) in a staggered arrangement at the bottom of the vessel. We have investigated  PMTs for their functionality and quantum efficiency at low temperature. The best result was obtained with  Hamamatsu PMTs (R5912-MOD) manufactured with Pt-underlay. A light but sufficiently strong mechanical support was  constructed to withstand the buoyant force ($\approx$1kN) acting on the PMTs in liquid argon.  The PMT glass was coated with a thin WLS layer of a transparent TPB-paraloid compound to increase the VUV light yield.  Fig. \ref{ArDMdetector} (right) shows the mounted foils and the PMTs under UV illumination. Many details can be found in a recent thesis \cite{Boccone}.

The ArDM detector was filled for the first time with 1t of LAr in 2009.  Several important parameters such as stable cryogenic operation in high LAr purity,   high scintillation light yield, and  detection of events down to energies of tens of keV's could be verified. The test was performed with a partial light readout assembly consisting of half of the PMTs, no electric field and no charge readout. The LAr purity was found to be constant over the measurement time of three weeks by monitoring the decay time $\tau_2$ of the slow component of the light signal \cite{amsler2}. 

\begin{figure}[htb]
\parbox{160mm}{\mbox{
\includegraphics[width=160mm]{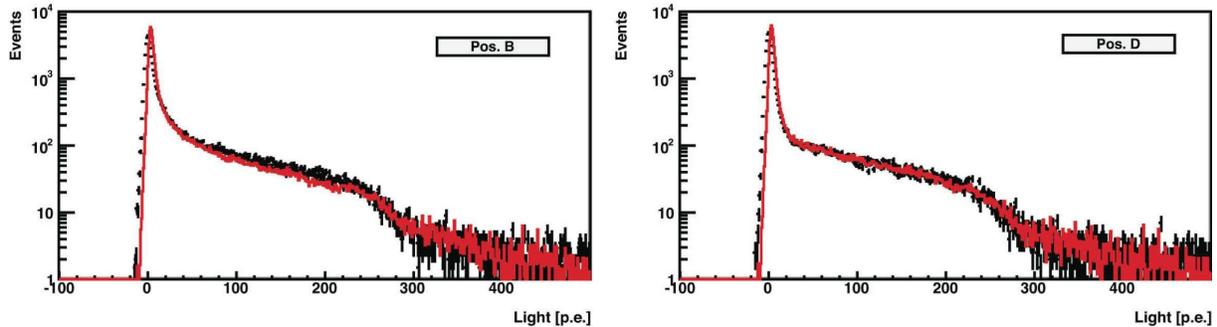}
}\centering}\hfill
\caption[]{\it Light yield  in the 1t  LAr detector for 511 keV $\gamma$'s (in photoelectrons, p.e.) at  two positions  of the $^{22}$Na- source. The measurements are in black, the simulated data in red (from ref. \cite{amsler2}).
\label{fig:DataMCComp}}
\end{figure}

The measurements were done with external sources such as $^{22}$Na, delivering positrons (annihilating into two 511 keV $\gamma$'s) and monochromatic 1275\,keV $\gamma
$'s. The light yield produced by one of the 511\,keV $\gamma$'s, following (multiple) Compton scattering, was measured by 
triggering with a 4" Na(Tl) crystal on the second 511\,keV emitted in the opposite direction, and on the 1275\,keV $\gamma$.   Fig. \ref{fig:DataMCComp} shows the light yield distributions 
for the source located at two different vertical distances from the photomultiplier array (black bars) \cite{amsler2}. The light yield (in photoelectrons, p.e.)  was  calculated by Monte Carlo simulation as a function of  energy deposit, 
and the distributions (red bars) compared to the measured ones. Good agreement was found with an average light yield of 
typically 0.4 p.e./keV, which is roughly half of the yield that would be obtained with a 
completed detector (14 PMTs). Note that the light yield for nuclear recoils is lower due to quenching, typically 30\% of that for electrons in the few 10 keV range, see Fig. \ref{fig:Leffstat} below.  We are 
therefore confident to be able to reach our  goal of  30 keV threshold in 1t  of LAr  for WIMP detection.  

\section{LAr response to neutrons}
\begin{figure}[htb]
\parbox{80mm}{\mbox{
\includegraphics[width=60mm]{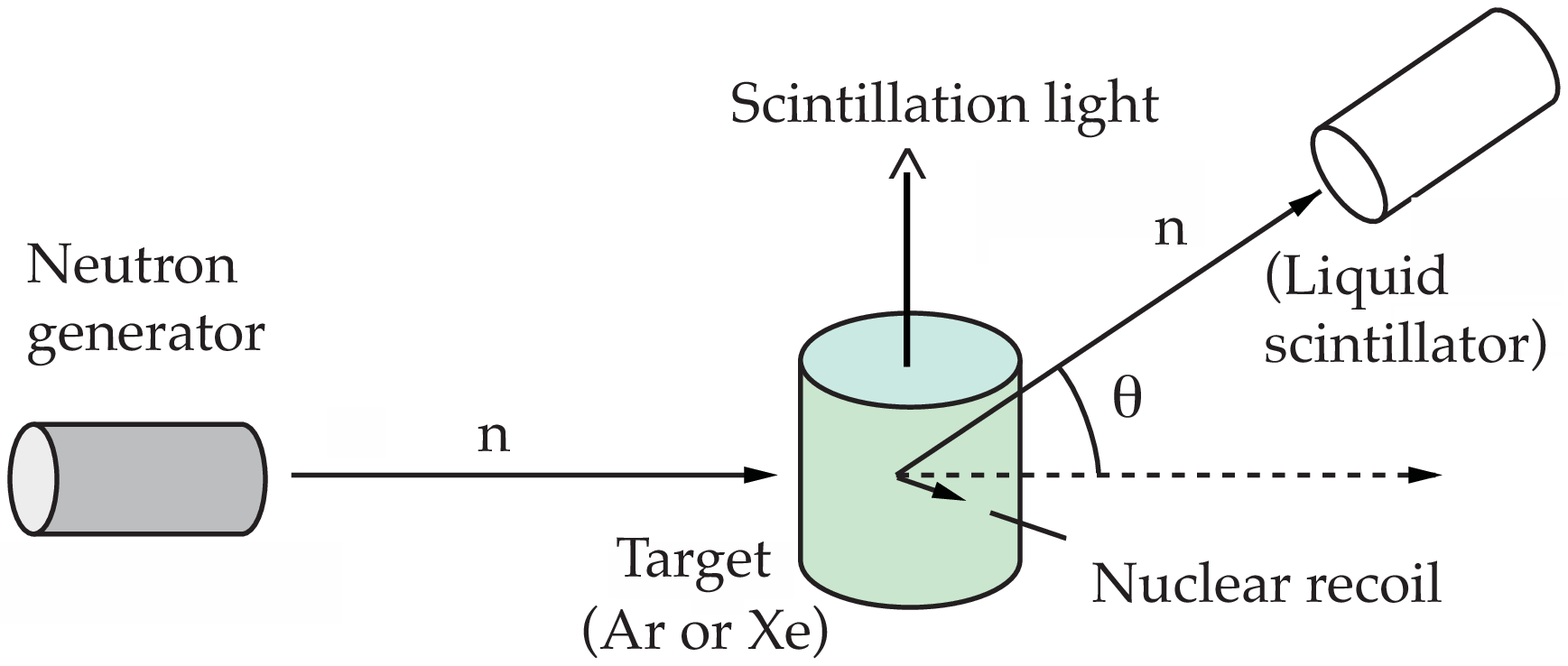}
}\centering}\hfill
\parbox{80mm}{\mbox{
\includegraphics[width=50mm]{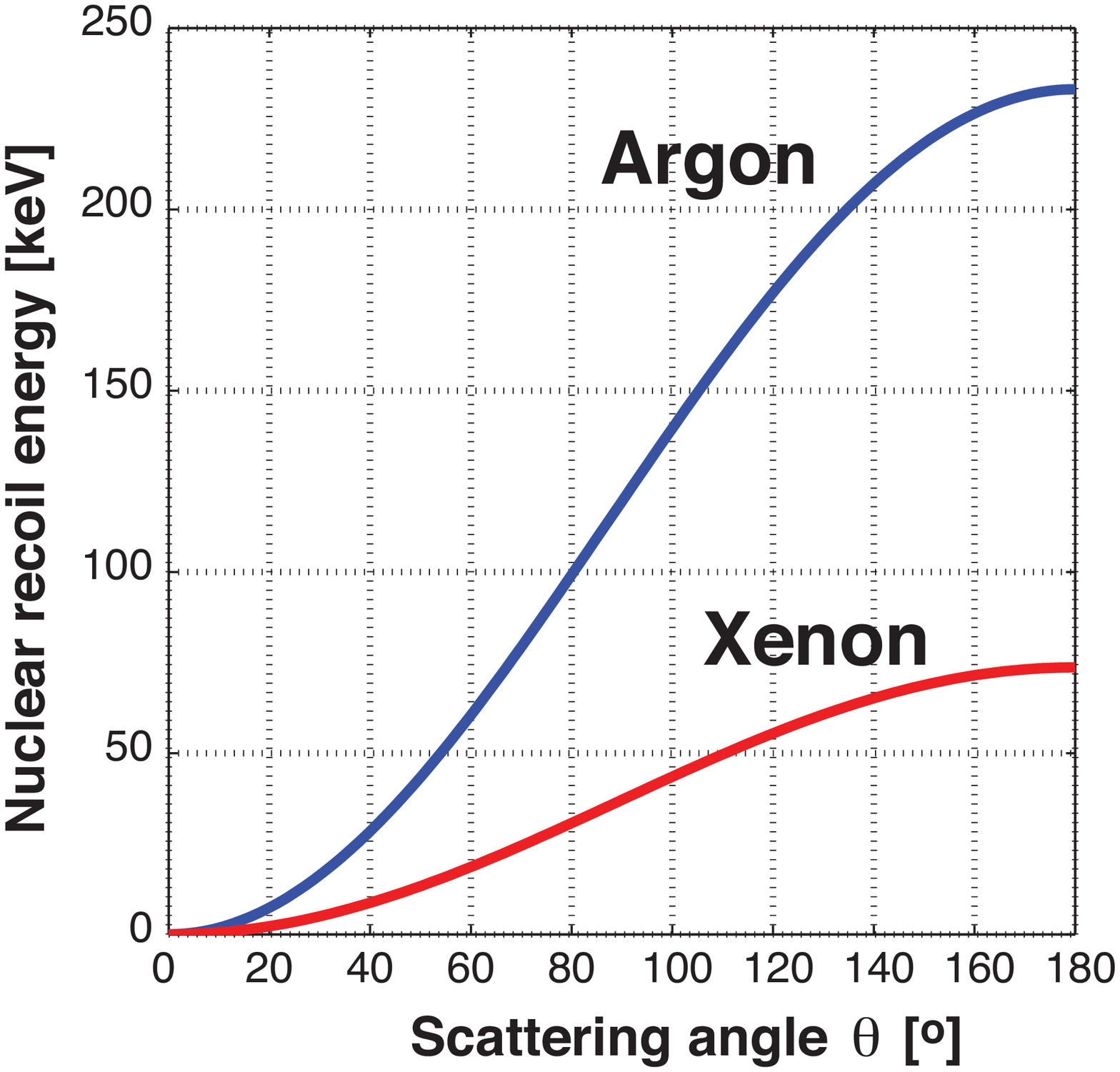}
}\centering} \caption[]{\it Left: Principle of a scattering experiment with monoenergetic neutrons. 
Right: nuclear recoil energies in argon and xenon for neutron energies of 2.45 MeV.
\label{Principle}}
\end{figure}

The  light yield of nuclear recoils in LAr   are poorly known, especially below 50 keV (see e.g. ref. \cite{mckin}). A good way to calibrate the light output and to study the response of LAr or LXe to nuclear recoils is to scatter a beam of monoenergetic neutrons of energy $T_n$ on argon or xenon nuclei and to measure the light yield as a function of scattering angle $\theta$, from which the recoil energy $T_r$ can be calculated according to the formula

\begin{equation}
T_r= \frac{2T_n \left[1+A-{\rm cos}^{2}\theta - \cos\theta\sqrt{A^{2}+{\rm cos}^{2}\theta-1}\right]}{(1+A)^{2}}\simeq \frac{2T_nA}{(1+A)^{2}}\left(1-\cos\theta\right),
\end{equation}
where $A$ is the atomic mass ($A \gg 1$). The method is illustrated in fig.~\ref{Principle} (left) and the energy deposits are plotted in Fig. \ref{Principle} (right) as a function of scattering angle for $T_n$ = 2.45 MeV incident neutrons.

In fact, neutrons (mainly from cosmic ray induced spallation) are a serious source of background in WIMP searches, due to the huge cross sections for neutron-nucleus scattering, many orders of magnitude larger than the ones for WIMP-nucleus scattering. This background contribution can be reduced by operating underground and by shielding the detector with low-$Z$ materials. However, the large neutron-nucleus cross section also means that the mean free path of neutrons is of the order of a few cm, leading to multiple scattering, quite in contrast to WIMPs which interact only once in the target. This gives an additional way to suppress neutron induced background, provided that multiple scattering can be detected from the enhanced light yield, and/or by identifying the multiple interaction vertices in the detector, e.g. with a charge detection device. 

\begin{figure}[htb]
\parbox{160mm}{\mbox{
\includegraphics[width=130mm]{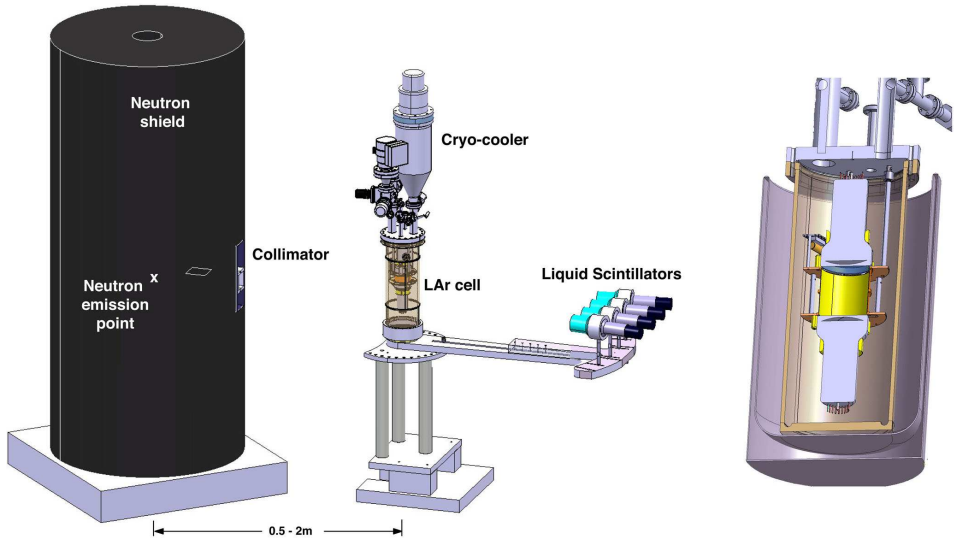}
}\centering}\hfill
\caption[]{\it Sketch of the neutron generator, the LAr  cell, the  cryocooler and the four liquid scintillators (LSCs) to detect scattered neutrons. The right 
picture shows  the LAr cell  with its 2 PMTs on top and bottom of the LAr volume.
\label{fig:ZUNFsetup}}
\end{figure}

We have therefore purchased a monoenergetic neutron source  \cite{nsd} based on the reaction $dd\to ^3$He $n$, and are setting up a scattering experiment with collimated 2.45 MeV neutrons. The target  is a  small  ($<$1$\ell$) test cell  (77 mm high and 74 mm in diameter) and liquid scintillation counters (LSC, EJ301 from SCIONIX) detect the scattered neutrons in coincidence as a function of scattering angle  (Fig.\,\ref{fig:ZUNFsetup}). To reduce the measurement time we use four LSCs to cover various  angles in parallel. 

\begin{figure}[htb]
\parbox{80mm}{\mbox{
\includegraphics[width=65mm]{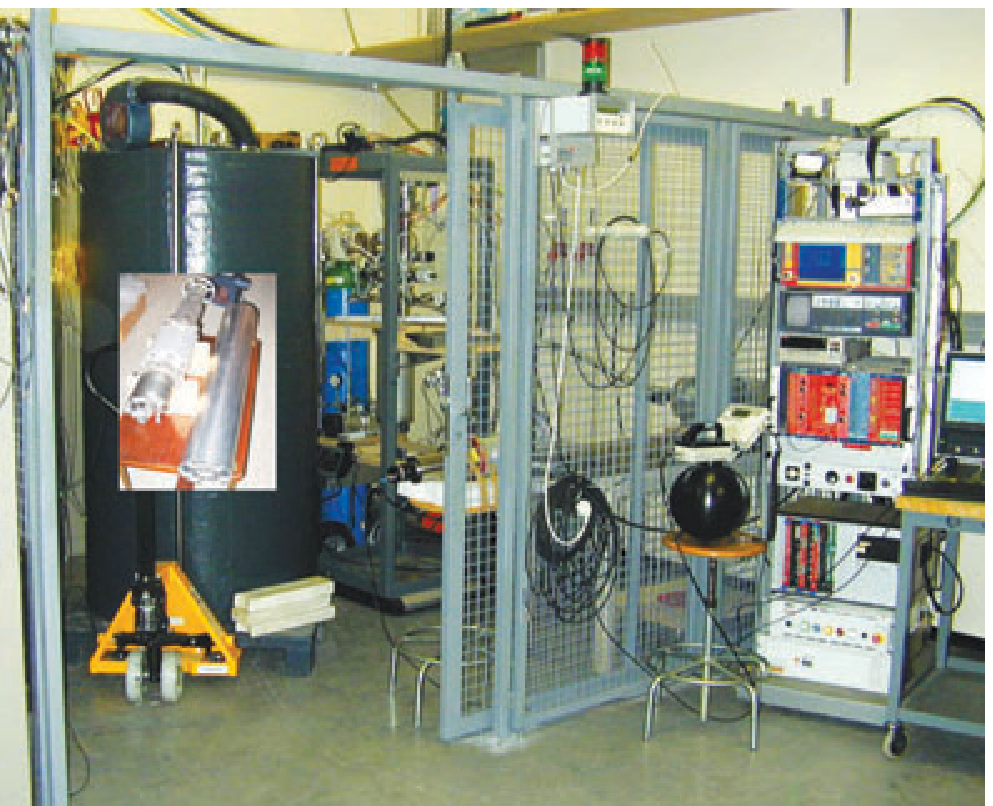}
}\centering}\hfill
\parbox{80mm}{\mbox{
\includegraphics[width=80mm]{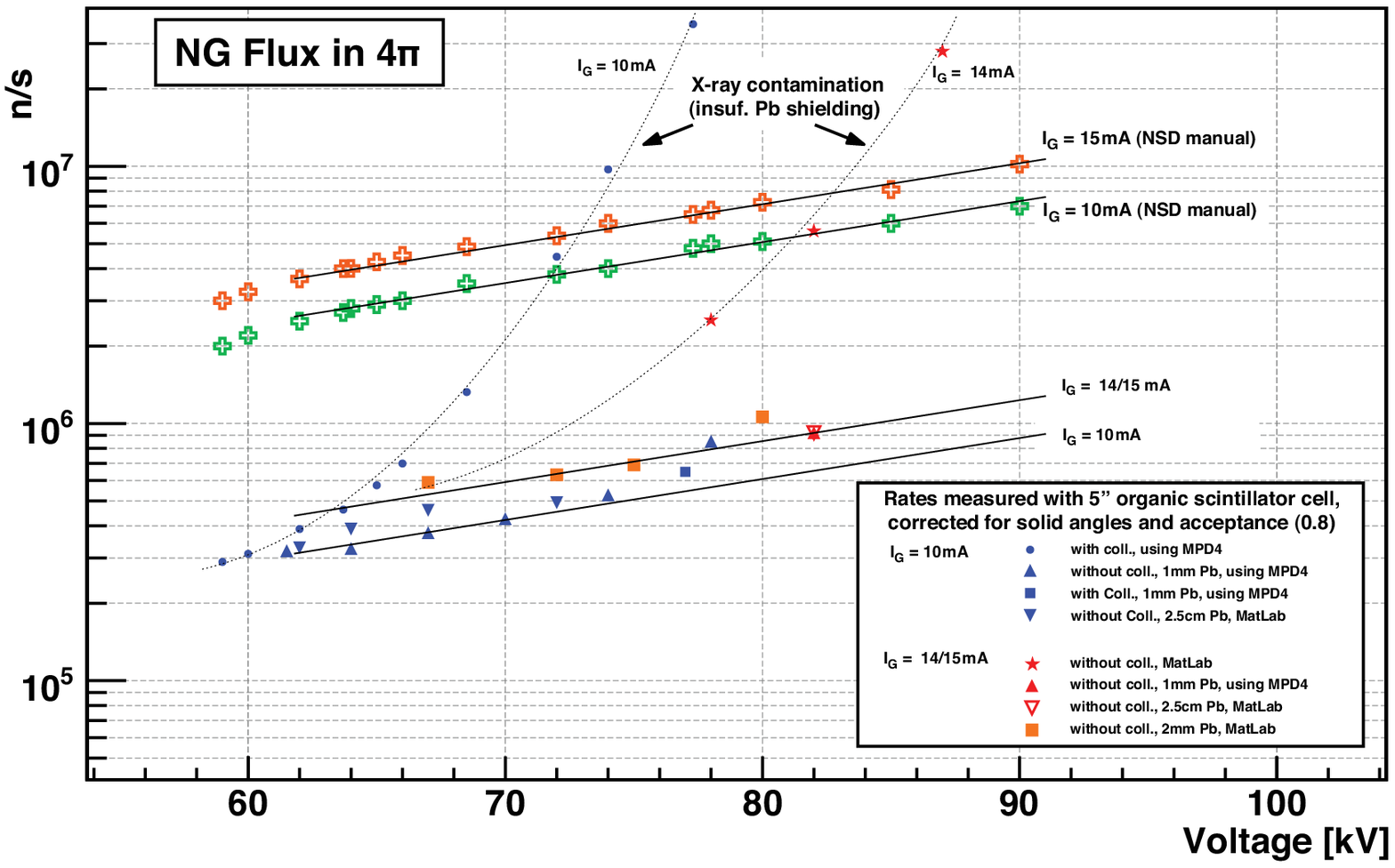}
}\centering} \caption[]{\it Left:  2.45 MeV neutron fusion generator  inside its safety fence in our laboratory at CERN. The source itself is shown in the inset. Visible are also the X/$\gamma$-ray monitor (top white box), the neutron radiation detector (black sphere) and the electronic rack (right) with the measuring electronics and the HV control units. 
Right: Neutron generator flux into 4$\pi$ (see text).  Straight lines are drawn to guide the eye.
\label{Generator}}
\end{figure}

The fusion chamber is surrounded by a 90 cm diameter shield of borated polyester and the experiment confined within a radiation controlled fence in our laboratory at CERN  (Fig. \ref{Generator}, left). Residual radiation (mainly  from scattered neutrons and X-rays)  is well below the authorized limit of 2.5\,$\mu$Sv/h. The neutrons are collimated through a polyethylene orifice within roughly 1\% $\times$ 4$\pi$ sr. The neutron flux  (up to 5$\times$10$^{6}$ n/s in 4$\pi$, according to specifications) is controlled through the applied high voltage and discharge current. 

The neutron flux was measured with a 5'' LSC located at the collimator exit. The  polyethylene collimator was surrounded by a 2 mm thick lead box against X-rays, inserted into the polyester shielding. According to NSD-Fusion \cite{nsd}, the highest possible voltage and current are 100 kV, resp. 15 mA, corresponding to a flux of $10^{7}$n/s into 4$\pi$. The  flux increases proportionally to the current and to the voltage V$^{2.8}$. Measurements were made with and without polyethylene collimator and the rates corrected for the solid angle, assuming a detection efficiency of the LSC of  80\%. A NIM MPD-4 module was used to discriminate between neutrons and X-rays. Figure  \ref{Generator} (right) shows the measured neutron intensity (blue,  red and orange dots) compared with the values specified by NSD-Fusion (green crosses for 10mA and orange crosses for 15mA \cite{nsd}). Insufficient lead shielding leads to a strong X-ray contamination.  The measured flux (red dots) was roughly one order magnitude smaller than anticipated, with a maximum of $10^{6}$ n/s. 

\begin{figure}[htb]
\parbox{80mm}{\mbox{
\includegraphics[width=60mm]{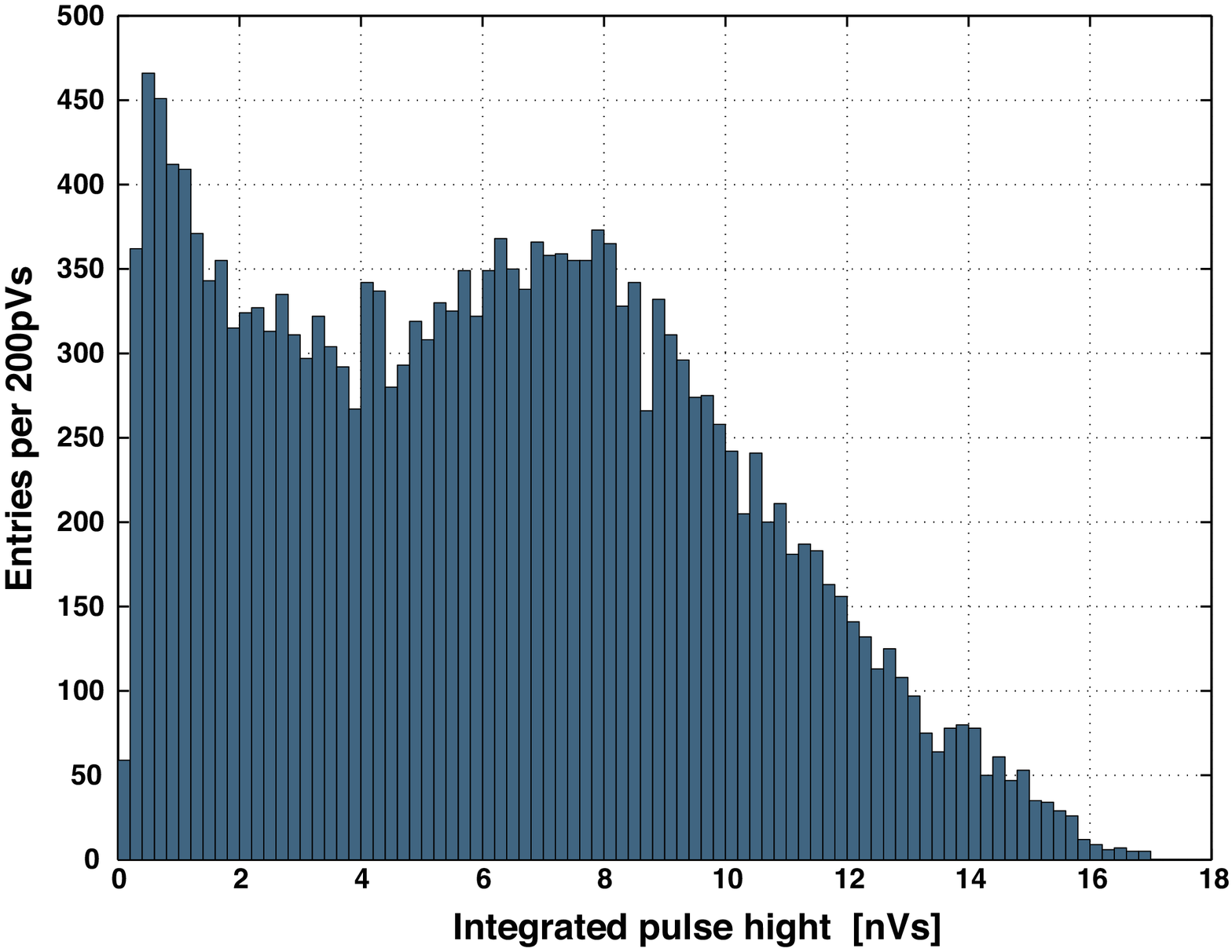}
}\centering}\hfill
\parbox{80mm}{\mbox{
\includegraphics[width=50mm]{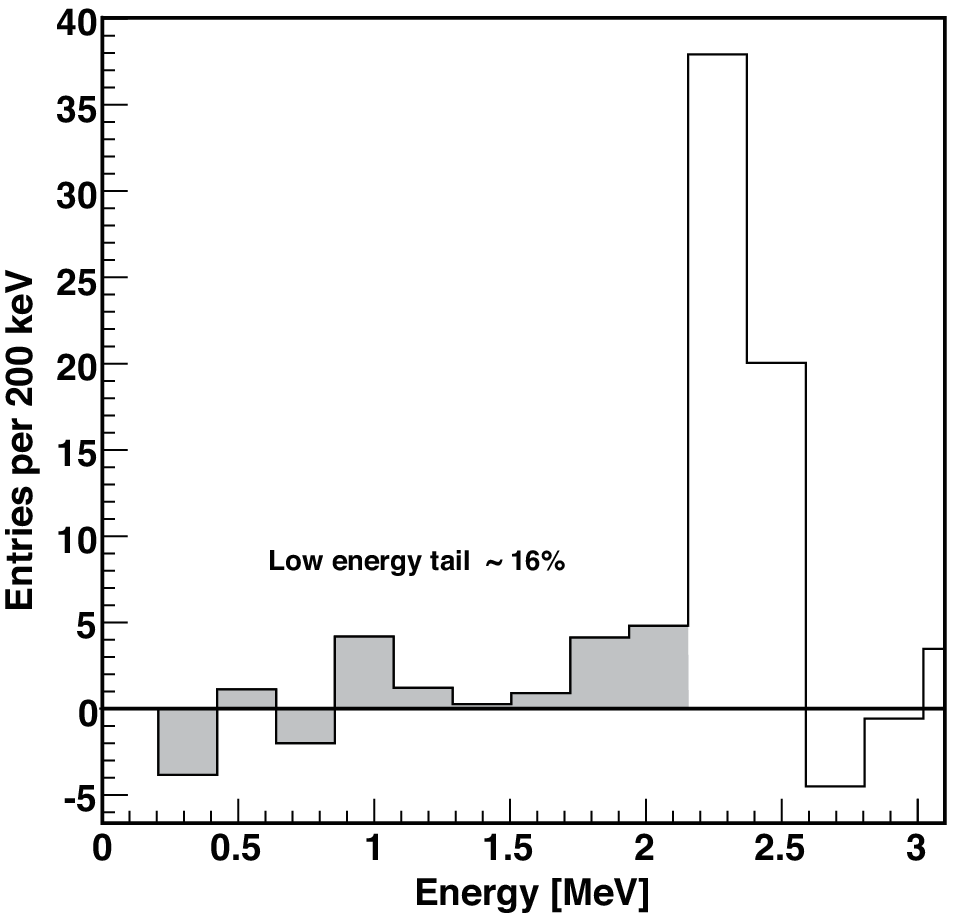}
}\centering} \caption[]{\it Left: Pulse height distribution induced by neutrons in the LSC at the exit of the collimator. Right: Incident neutron energy spectrum obtained after unfolding the response of the LSC.
\label{fig:unfold}}
\end{figure}

The neutron  energy distribution (smeared by  collimator scattering)  must be known to measure the light yield from LAr accurately.  Fig. \ref{fig:unfold} (left) shows the neutron energy distribution measured by a 5'' LSC placed at the exit of the collimator. Ideally the spectrum should be flat (without collimator scattering, with infinite resolution and single $n$-scattering). The spectrum of Fig. \ref{fig:unfold} (right) was obtained by unfolding the response of the LSC to neutrons from a radioactive source. The response was obtained with an AmBe-source, by measuring the neutron energy through the time-of-flight between the source and the neutron counter over a distance of 1 m. The start time was determined from the 4.4 MeV $\gamma$ detected in a BGO crystal located close to the AmBe-source. 

\begin{figure}[htb]
\parbox{160mm}{\mbox{
\includegraphics[width=130mm]{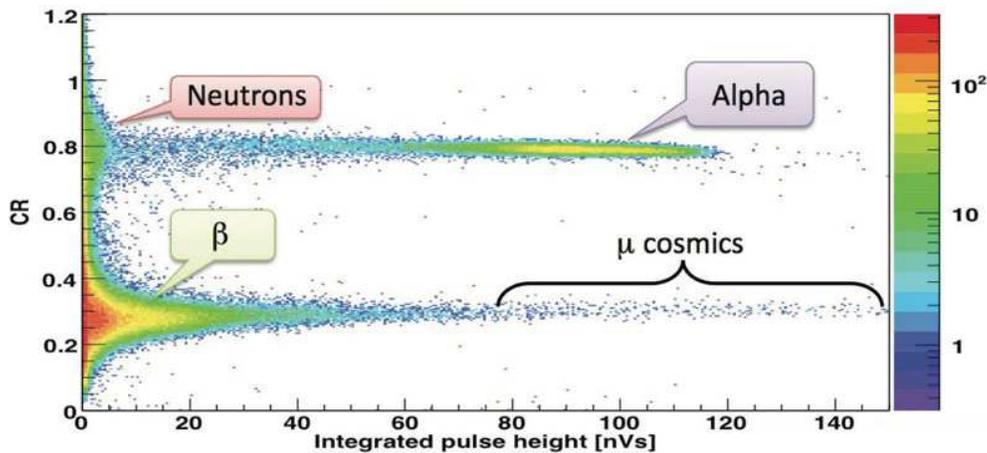}
}\centering}\hfill
\caption[]{\it Component ratio $CR$ in LAr for 2.45 MeV neutrons, 5.3 MeV $\alpha$'s and 1.3 MeV electrons and cosmic muons.
\label{CRNeutrons}}
\end{figure}

We have started to measure the scintillation response of LAr to nuclear recoils with the LAr cell shown in Fig. \ref{fig:ZUNFsetup} (right). Wavelength shifting reflectors (Tetraphenyl-Butadiene, TPB,  on Tyvek foils) were mounted on the  inner walls of the cell to convert the 128 nm light into 400 nm. The cell was read out by two Hamamatsu R6091-01MOD PMTs with Pt-underlay for cryogenic operation. An internal $^{210}$Pb-source emitted 5.3 MeV $\alpha$'s and up to 1.2 MeV electrons. We recall that the component ratio $CR$ is the ratio of integrated light yield during the first 50 ns to the total light yield. Thus a high $CR$ corresponds to the emission of light with mainly the fast component. As mentioned above, heavily ionizing particles such as $\alpha$'s or nuclear recoils lead to a large $CR$ value. Fig. \ref{CRNeutrons} shows the component ratio $CR$ from one of our first measurements of argon luminescence with the fusion generator. A clear  contribution from neutron induced nuclear recoils  is  observed at $CR\sim 0.8$. 

A first measurement of the scintillation efficiency for nuclear recoils relative to electrons was performed with a 5''  LSC at  $65^{\circ}$ from the incident beam direction, at a distance of 50 cm from the LAr cell.  The time-of-flight between the LAr cell and the LSC could be determined off-line and used to remove background, e.g. from multiple neutron scattering in the cell. The reference time was determined with a Na-source located at equal distance from the LAr-cell and the LSC, and using the two back-to-back 511 keV $\gamma$'s.  At $65^{\circ}$ the argon recoil energy  is 69 keV with 2.45 MeV incident neutrons. This corresponds to a time-of-flight of 23 ns for 2.38 MeV neutrons flying to the LSC. Good pulse shape separation between proton and background ($\gamma$-induced electron recoils) could be achieved with  the LSC by analogue pulse shape discrimination. 

\begin{figure}[htb]
\parbox{80mm}{\mbox{
\includegraphics[width=70mm]{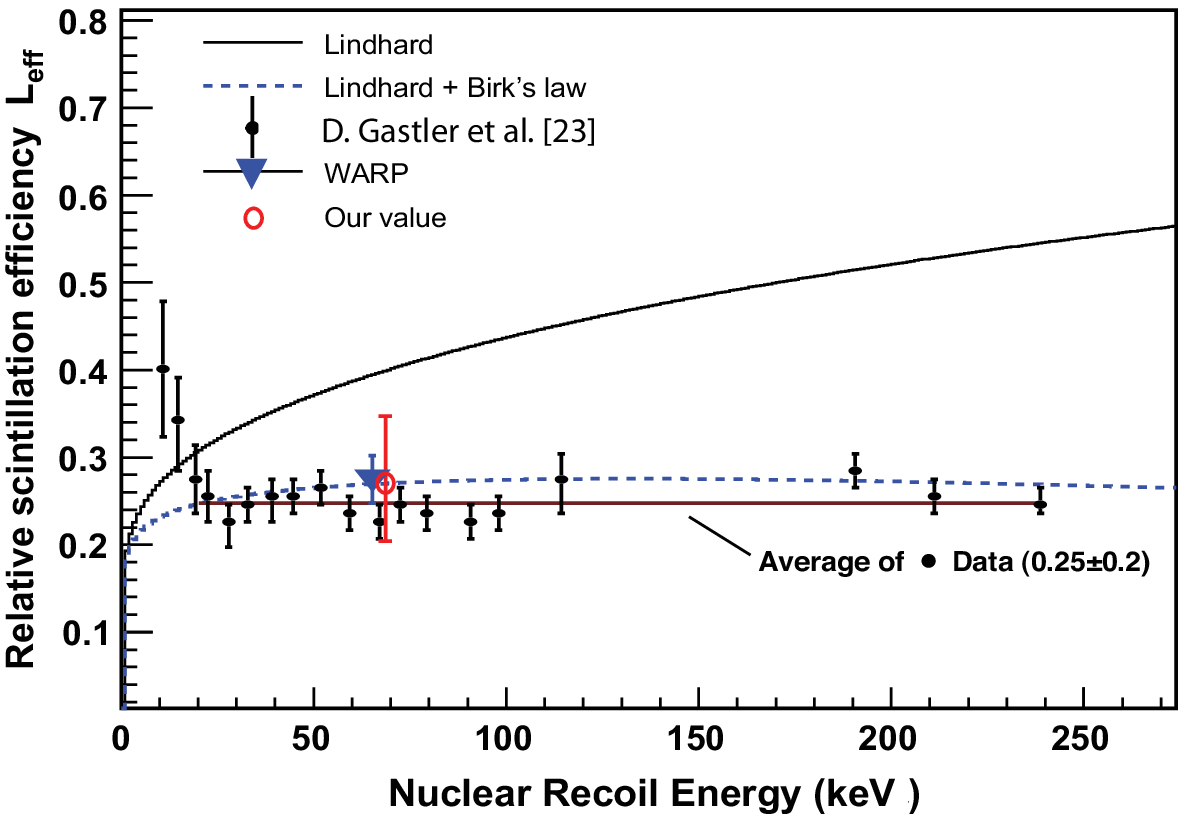}
}\centering}\hfill
\parbox{80mm}{\mbox{
\includegraphics[width=70mm]{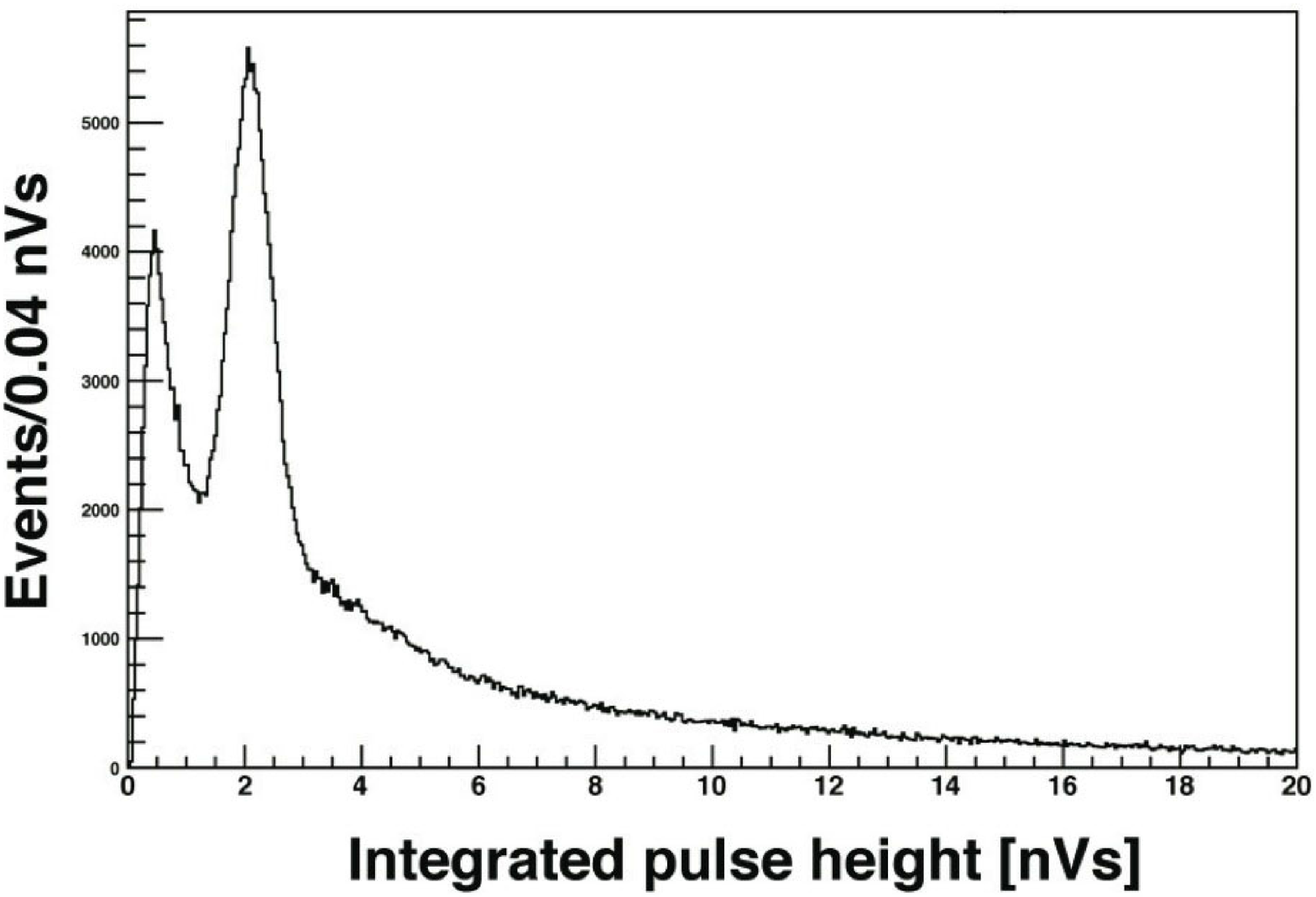}
}\centering} \caption[]{\it Left:  Measurement of the relative scintillation efficiency in LAr as a function of recoil energy (from ref. \cite{mckin}). Our preliminary measurement is shown by the red data point. Right: Energy spectrum from  Am-decay measured with the refurbished LAr cell, showing the 60 keV line.
\label{fig:Leffstat}}
\end{figure}

Several cuts were applied in the offline analysis. For example, we required the component ratio $CR$ to be larger than 0.6 (see Fig. \ref{CRNeutrons}). We also rejected events close to the PMT windows by comparing the signals from the two PMTs.  A time-of-flight window between 18 and 31 ns was selected.  We obtained a scintillation efficiency  of 0.27 (with about 25\% error) in this first attempt. Fig. \ref{fig:Leffstat} (left) shows our data point compared with data from ref. \cite{mckin}.

Light yields of nuclear recoils are usually determined relative to electronic recoils. The electronic light yield is determined with various external $\gamma$-sources and also with a $^{83}$Kr$^{m}$ source which can be connected 
directly to the gaseous phase in our setup. 

To improve the light yield and the measurements with radioactive sources we upgraded our LAr cell to reduce the thickness of the  stainless steel vessel and the LAr volume. Tetratex foils (from Donaldson Membranes) were used instead of the Tyvek reflectors, and TPB-Paraloid coating was replaced by 0.08 mg$\cdot$cm$^{-2}$ of evaporated TPB. Fig. \ref{fig:Leffstat} (right) shows the spectrum of the 60 keV line from Am-decay. During data taking the mean life of the 
slow component was  much lower than the established value of 1.6 $\mu$s, due to the impurity of the LAr (see e.g. ref. \cite{amsler2}). By extrapolating the light yield to maximum purity  we could set a lower limit of 3.2 p.e./keV.

The setup has now been improved by the addition of  a cryocooler (Gif\-ford-Mc\-Mahon-type) moun\-ted on top of the LAr cell. A cooling serpentine is bonded to the temperature regulated cold head and provides the liquefaction of the recirculated gas. Gas purification is achieved by two cleaning cartridges. They reduce the O$_{2}$ and H$_{2}$O contamination below 20 ppm. A further upgrade of the LAr cell with a smaller active volume (and less surrounding material) to reduce multiple scattering and enhance the light yield is in progress.

\section{Conclusions}
The current experimental sensitivity on WIMPs is around 10$^{-8}$ pb for masses of $\sim$100 GeV. The goal of future experiments using solid state detectors or noble liquids such as LAr or LXe is to reach 10$^{-10}$ -- 10$^{-11}$ pb, a sensitivity within theoretical predictions from supersymmetry. The goal of the DARWIN consortium is to deliver the design for a LXe/LAr detector to become operational around 2016. Liquid argon technology is still in its infancy and intensive R\&D is ongoing, in particular to determine the light yield from heavily ionizing nuclear recoils. For  the first time a large 1t LAr argon prototype detector (ArDM) was operated on the surface with a light yield of 0.8 photoelectrons/keV electron equivalent. These are encouraging prospects to detect  WIMPs in LAr with a threshold below 50 keV.
\section*{Acknowledgments}
This work was performed together with the ArDM \cite{ArDM}  and DARWIN \cite{DARWIN} collaborations. Contributions from my collaborators of the Physik-Institut (Y. Allkofer, V. Boccone,  W.~Creus,  A. Ferella, P. Otyugova, C. Regenfus, J. Rochet, L. Scotto Lavina,  and M. Walter) are gratefully acknowledged.

\end{document}